\documentclass[preprint,11pt]{elsarticle}
\usepackage{amssymb}
\newtheorem {Theorem}                 {Theorem}         [section]

\newtheorem {lemma}        [Theorem]  {Lemma}
\newtheorem {Procedure}    [Theorem]  {Procedure}
\usepackage{pgfplots}
\usepackage{tikz}
\usetikzlibrary{shapes,arrows,backgrounds,mindmap}
\usepackage{titlesec}
\journal{arXiv}
\usetikzlibrary{shapes,arrows,backgrounds,mindmap}
\usepackage{subfigure}

\begin{document}
\begin{frontmatter}
\title{On testing single connectedness in directed graphs and some related problems}
\address{Department of Computer Science and Automation, Technische Universit\"at Ilmenau, $98693$ Ilmenau, Germany}
\author{Martin Dietzfelbinger}
\ead{martin.dietzfelbinger@tu-ilmenau.de} 
\author{Raed Jaberi} 
\ead{raed.jaberi@tu-ilmenau.de}

\begin{abstract}
Let $G=(V,E)$ be a directed graph with $n$ vertices and $m$ edges. The graph $G$ is called singly-connected if for each pair of vertices $v,w \in V$ there is at most one simple path from $v$ to $w$ in $G$. Buchsbaum and Carlisle (1993) gave an algorithm for testing whether $G$ is singly-connected in $O(n^{2})$ time. In this paper we describe a refined version of this algorithm with running time $O(s\cdot t+m)$, where $s$ and $t$ are the number of sources and sinks, respectively, in the reduced graph $G^{r}$ obtained by first contracting each strongly connected component of $G$ into one vertex and then eliminating vertices of indegree or outdegree $1$ by a contraction operation. Moreover, we show that the problem of finding a minimum cardinality edge subset $C\subseteq E$ (respectively, vertex subset $F\subseteq  V$) whose removal from $G$ leaves a singly-connected graph is NP-hard.  
\end{abstract}
\end{frontmatter}
\section{Introduction}
Let $G=(V,E)$ be a directed graph with $n$ vertices and $m$ edges. The graph $G$ is called \textit{singly-connected} if for every pair of vertices $v,w \in V$ there is at most one simple path from $v$ to $w$ in $G$. The problem of testing whether or not $G$ is singly-connected was introduced by Cormen et al. in \cite[Ex. 23.3-10]{CLR91} and \cite[Ex. 22.3-13]{CLRS09}. In \cite{BC93,BC92}, Buchsbaum and Carlisle presented an algorithm for solving this problem in $O(n^{2})$ time. Khuller described in \cite{K99} a similar approach for solving the same problem in $O(n^{2})$ time. Karlin \cite{K95}   (also see \cite{K00}) also presented a simple $O(n^{2})$ algorithm for solving the problem.   

Let $G=(V,E)$ be a directed graph. By $G^{c}=(V^{c},E^{c})$ we denote the directed graph which is obtained by contracting every strongly connected component of $G$ into a single vertex. The algorithms from \cite{BC93} and \cite{K99} are based on reducing the problem on $G$ to the same problem on the acyclic graph $G^{c}$. We use $G^{r}$ to denote the graph obtained from $G^{c}$ by eliminating all vertices of indegree or outdegree $1$ by contraction operations. A vertex $x$ of $G^{r}$ is called a source if its indegree is $0$ and a vertex $y$ of $G^{r}$ is called a sink if its outdegree is $0$. By $s$ and $t$ we denote the number of sources and sinks in $G^{r}$, respectively. In this paper we improve the running time of the algorithms from \cite{BC93},\cite{K99} from $O(n^{2})$ to $O(s\cdot t+m)$. We also give an example for a graph where $s\cdot t$ is much bigger than $m$. The question posed by Khuller \cite{K99} whether the problem of testing single connectedness in directed graphs is solvable in linear time remains open.

As mentioned in \cite{BC92}, it is clear that a singly-connected graph can be obtained from a directed graph which is not singly-connected by removing edges, but the property of singly-connectivity can be ruined by adding edges. We also show that the problem of finding a minimum cardinality edge subset $C\subseteq E$ (respectively, vertex subset $F\subseteq  V$) whose removal from $G$ leaves a singly-connected graph is NP-hard.

Papers \cite{BC93} and \cite{K99} show the existence of a procedure which has the following behavior: 
\begin{figure}[htbp]
\begin{Procedure}\label{algo:procSC}\rm\quad\\[-5ex]
\begin{tabbing}
\quad\quad\=\quad\=\quad\=\quad\=\quad\==\kill
\textbf{Input:} A directed graph $G=(V,E)$.\\
\textbf{Output:} \\
{\small }\> ``not singly-connected'' \\
\textbf{or}\\
{\small }\> ``Test whether $G^{c}$ is singly-connected''.
\end{tabbing}
\end{Procedure}
\end{figure} 

Procedure \ref{algo:procSC} without testing of $G^{c}$ has running time $O(m)$. Paper \cite{BC93} shows that the acyclic graph $G^{c}$ is singly-connected if and only if for every vertex $w\in V^{c}$ the vertex set of the DFS tree rooted at $w$ has neither cross edges nor forward edges. Testing single connectedness of $G^{c}$ in this way takes $O(n^{2})$ time \cite{BC93,K99} since there are $n$ calls to DFS, one for each vertex of $G^{c}$, and one can stop as soon as a forward edge or cross edge appears. The procedure leads to the situation that we only have to consider acyclic graphs. 
We give another reduction to be applied after Procedure \ref{algo:procSC} (which can be carried out in time $O(m)$) to a reduced graph $G^{r}$, so that $G$ is singly-connected if and only if $G^{r}$ is. In $G^{r}$ no vertex has indegree or outdegree $1$. 
\section{Improved handling of acyclic graphs}
\begin{figure}[htbp]
    \centering
    \subfigure[$G^{c}$]{
    \begin{tikzpicture}
        \tikzstyle{every node}=[color=black,draw,circle,minimum size=0.7cm]
        \node  (v0) at (0,3) {$0$};
        \node  (v1) at (1,1.8) {$1$};
        \node  (v2) at (1,0) {$2$};
        \node  (v3) at (2,3) {$3$};
        \node  (v4) at (-0.5,-1) {$4$};
        \node  (v5) at (2,-1) {$5$};
        \node  (v6) at (3,1) {$6$};
        \node  (v7) at (-1.2,1) {$7$};

       \begin{scope}
            \tikzstyle{every node}=[auto=right]
            \draw[-triangle 45] (v0) to   (v1);
            \draw[-triangle 45] (v1) to   (v2);
            \draw[-triangle 45] (v3) to   (v1);
            \draw[-triangle 45] (v2) to   (v4);
            \draw[-triangle 45] (v3) to   (v6);
            \draw[-triangle 45] (v2) to   (v5);
            \draw[-triangle 45] (v6) to   (v5);
            \draw[-triangle 45] (v7) to   (v4);
            \draw[-triangle 45] (v0) to   (v7);
            \draw[-triangle 45] (v1) to   (v7);
        \end{scope}
    \end{tikzpicture}} \qquad\qquad
    \subfigure[]{
    \begin{tikzpicture}
        \tikzstyle{every node}     =[color=black,draw,circle,minimum size=0.7cm]
        \node  (v0) at (0,3) {$0$};
        \node  (v1) at (1,1.8) {$1$};
        \node  (v3) at (2,3) {$3$};
        \node  (v4) at (-0.5,-1) {$4$};
        \node  (v5) at (2,-1) {$5$};
        \node  (v6) at (3,1) {$6$};
        \node  (v7) at (-1.2,1) {$7$};

       \begin{scope}
            \tikzstyle{every node}=[auto=right]
            \draw[-triangle 45] (v0) to   (v1);
            \draw[-triangle 45] (v3) to   (v1);
            \draw[-triangle 45] (v1) to   (v4);
            \draw[-triangle 45] (v3) to   (v6);
            \draw[-triangle 45] (v1) to   (v5);
            \draw[-triangle 45] (v6) to   (v5);
            \draw[-triangle 45] (v7) to   (v4);
            \draw[-triangle 45] (v0) to   (v7);
            \draw[-triangle 45] (v1) to   (v7);
         
        \end{scope}
    \end{tikzpicture}}
   \caption{(a) $G=(V,E)$, vertex $2$ has indegree $1$. (b) The remaining graph after contracting edge $(1,2)$.}
    \label{fig:FigForSchrinkOperation}
\end{figure}
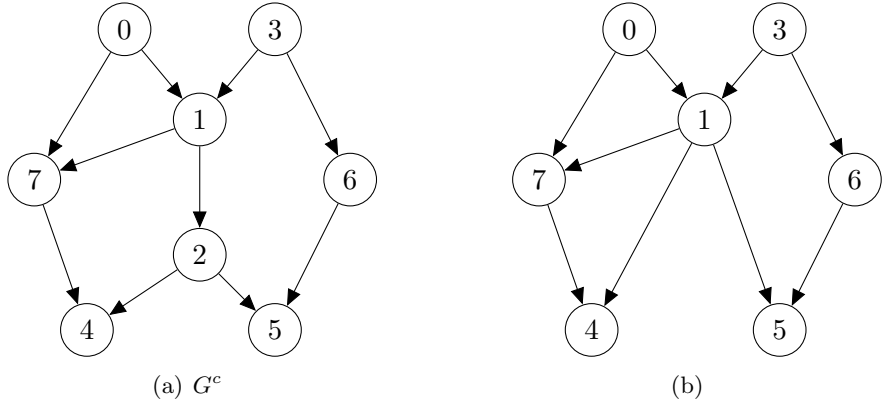

Assume that the input graph $G=(V,E)$ is acyclic. We propose two types of improvement over the algorithms in \cite{BC93,K99} for acyclic graphs:
\begin{itemize}
\item[1.] eliminating vertices with indegree or outdegree $1$.
\item[2.] starting DFS only from sources.
\end{itemize}
In the first step, called preprocessing step, we modify $G$ as follows. We consider the vertices in bottom-up order. For each vertex $v\in V$ with indegree $1$ we shrink $u,v$, where $(u,v) \in E$, by replacing each edge $(v,w)\in E$ by $(u,w)$ and removing the vertex $v$ and the edge $(u,v)$ from $G $ (see Figure \ref{fig:FigForSchrinkOperation} for one such contraction). 
Of course, if $v$ has outdegree $0$, it can simply be deleted. Let $G'$ be the resulting graph. To $G'$ we apply a similar procedure top down to eliminate all nodes of outdegree $1$. The resulting graph is called $G^{r}$.
\begin{lemma}\label{def:stepOneInLinearTimeAndPreservingSC}

\item[(a)] $G^{r}$ can be computed in time $O(m)$.
\item[(b)] $G$ is singly-connected if and only if $G^{r}$ is singly-connected.
\end{lemma}
\emph{Proof} 
(a) We represent the acyclic graph $G$ by adjacency lists in which each vertex $v\in V$ has a circular doubly linked list $L_{v}$ containing all the successors of $v$ in $G$. Eliminate all vertices of indegree $1$, as follows:
\begin{itemize}
\item[(i)] Treat vertices $v$ ``bottom up'' (in the partial order given by $G$) as follows: If $v$ has  indegree $1$, merge $v$ with its predecessor $u$ by linking $L_{v}$ into $L_{u}$ in constant time and deleting $v$. It is clear this can be done in time $O(m)$. (Actually, the merging itself takes time $O(n)$, only finding vertices with indegree $1$ needs $O(m)$ time.)
 \item[(ii)] Eliminate all vertices of outdegree $1$. This is done exactly as in (i), just using the reversed graph of $G'$.
\end{itemize}
(b) Each step preserves the property of single connectedness. \qed\\
Note that if multiple edges arise in the preprocessing step, then $G$ is not singly-connected. Check once at the very end if $G^{r}$ has multiple edges. If so, return `` not singly-connected''.

From here on we only consider the reduced graph $G^{r}$, in which all non-sources have indegree at least $2$ and all non-sinks have outdegree at least $2$.
 In the second step we perform a DFS only for the sources of $G^{r}$. The correctness of this step is based on the following lemma.
\begin{lemma}
Let $G=(V,E)$ be a directed acyclic graph such that $G$ does not have any multiple edges. Then $G$ is singly-connected if and only if for every source $v \in V$ the vertex set of the DFS tree $T_{v}$ with root $v$ has only tree edges. 
\end{lemma}
\emph{Proof}
``$\Leftarrow$'': Assume that $G$ is not singly-connected. Then by definition, there are two vertices $v,w\in V$ such that there exist at least two simple paths $p_{1},p_{2}$ from $v$ to $w$ in $G$. Then there is a vertex $u$ that lies on both paths $p_{1},p_{2}$ and has two different incoming   edges in $p_{1} \cup p_{2}$. There is some source $s$ such that there is a path from $s$ to $v$ in $G$. Hence the vertex $u$ and all the vertices on $p_{1} \cup p_{2}$ are in $T_{s}$.
Of course, it is impossible that both edges entering $u$ are tree edges. \\
``$\Rightarrow$'': Any forward or cross edge in a DFS from any vertex immediately proves that $G$  is not singly-connected (see \cite{BC93}).\qed
\begin{Theorem}\label{def:RunningTimeOfModifiedAlgorithm}
It can be tested in $O(s\cdot t+m)$ time if $G^{r}$ is singly-connected.
\end{Theorem}
\emph{Proof} As we have ssen, the transformation from $G$ to $G^{r}$ takes time $O(m)$. In $G^{r}$, each DFS tree $T$ has at most $2t-1$ vertices since each vertex which is not a leaf in $T$ has outdegree at least $2$ and all leaves of $T$ are sinks so there cannot be more than $t
$ leaves. Each DFS requires $\Theta(t)$ time because the algorithm stops as soon as a cross edge or a forward edge appears. The total number of DFS-trees is at most $s$. Therefore, the total running time is $O(s\cdot t+m)$.\qed\\
One may ask whether there are singly-connected graphs in which $s\cdot t$ is much bigger than $m$ and in which our algorithm takes much longer than $O(m)$ time. Actually, a well known graph family, the butterfly graphs (or FFT graphs) $B_{d}$ (see, e.g., \cite{T92}), give an example. They have $n=2^{d}(d+1)$ vertices and $m=2^{d+1}d $ edges, are singly-connected, and we have $s=t=2^{d}$ and hence $s\cdot t=t^{2}=2^{2d}\gg m $. 

\begin{figure}[htbp]
    \centering
    \begin{tikzpicture} 
        \tikzstyle{every node}=[color=black,draw,circle,minimum size=0.7cm]
        \node (v0) at (0,0) {};
        \node (v1) at (1.5,0) {};
        \node (v2) at (3,0) {};
        \node (v3) at (4.5,0) {};
        \node (v4) at (6,0) {};
        \node (v5) at (7.5,0) {};
        \node (v6) at (9,0) {};
        \node (v7) at (10.5,0) {};
        \node (v8) at (0,2) {};
        \node (v9) at (1.5,2) {};
        \node (v10) at (3,2) {};
        \node (v11) at (4.5,2) {};
        \node (v12) at (6,2) {};
        \node (v13) at (7.5,2) {};
        \node (v14) at (9,2) {};
        \node (v15) at (10.5,2) {};
         \node (v16) at (0,4) {};
        \node (v17) at (1.5,4) {};
        \node (v18) at (3,4) {};
        \node (v19) at (4.5,4) {};
        \node (v20) at (6,4) {};
        \node (v21) at (7.5,4) {};
        \node (v22) at (9,4) {};
        \node (v23) at (10.5,4) {};
        \node (v24) at (0,6) {};
        \node (v25) at (1.5,6) {};
        \node (v26) at (3,6) {};
        \node (v27) at (4.5,6) {};
        \node (v28) at (6,6) {};
        \node (v29) at (7.5,6) {};
        \node (v30) at (9,6) {};
        \node (v31) at (10.5,6) {};
       \begin{scope}
            \tikzstyle{every node}=[auto=right]
            \draw [-triangle 45]   (v12) to(v0);
            \draw [-triangle 45]   (v30) to(v23);
            \draw [-triangle 45]   (v31)to(v22);
            \draw [-triangle 45]   (v28)to(v21);
            \draw [-triangle 45]   (v29)to(v20);
            \draw [-triangle 45]   (v26)to(v19);
            \draw [-triangle 45]   (v27)to(v18);
            \draw [-triangle 45]   (v24)to(v17);
            \draw [-triangle 45]   (v25)to(v16);
            \draw [-triangle 45]   (v31)to(v23);
            \draw [-triangle 45]   (v30)to(v22);
            \draw [-triangle 45]   (v29)to(v21);
            \draw [-triangle 45]   (v28)to(v20);
            \draw [-triangle 45]   (v27)to(v19);
            \draw [-triangle 45]   (v26)to(v18);
            \draw [-triangle 45]   (v25)to(v17);
            \draw [-triangle 45]   (v24)to(v16);
            \draw [-triangle 45]   (v23)to(v15);
            \draw [-triangle 45]   (v22)to(v14);
            \draw [-triangle 45]   (v21)to(v13);
            \draw [-triangle 45]   (v20)to(v12);
            \draw [-triangle 45]   (v19)to(v11);
            \draw [-triangle 45]  (v18)to(v10);
            \draw [-triangle 45]   (v17)to(v9);
            \draw [-triangle 45]   (v16)to(v8);
            \draw [-triangle 45]   (v15)to(v7);
            \draw [-triangle 45]   (v14)to(v6);
            \draw [-triangle 45]  (v13)to(v5) ;
            \draw [-triangle 45]  (v12)to(v4);
            \draw [-triangle 45]   (v11)to(v3);
            \draw [-triangle 45]  (v10)to(v2) ;
            \draw [-triangle 45]   (v9)to(v1);
            \draw [-triangle 45]   (v8)to(v0);
            \draw [-triangle 45]   (v21)to(v15);
            \draw [-triangle 45]   (v20)to(v14);
            \draw [-triangle 45]   (v23)to(v13);
            \draw [-triangle 45]   (v22)to(v12);
            \draw [-triangle 45]  (v17)to(v11) ;
            \draw [-triangle 45]   (v16)to(v10);
            \draw [-triangle 45]   (v19)to(v9);
            \draw [-triangle 45]   (v18)to(v8);
            \draw [-triangle 45]   (v11)to(v7);
            \draw [-triangle 45]   (v10)to(v6);
            \draw [-triangle 45]   (v9)to(v5);
            \draw [-triangle 45]   (v8)to(v4);
            \draw [-triangle 45]   (v15)to(v3);
            \draw [-triangle 45]   (v14)to(v2);
            \draw [-triangle 45]   (v13)to(v1);  
        \end{scope}
    \end{tikzpicture} 
   \caption{ Butterfly graph $B_{3}$.}
    \label{fig:ExpButterflyGraph}
\end{figure}
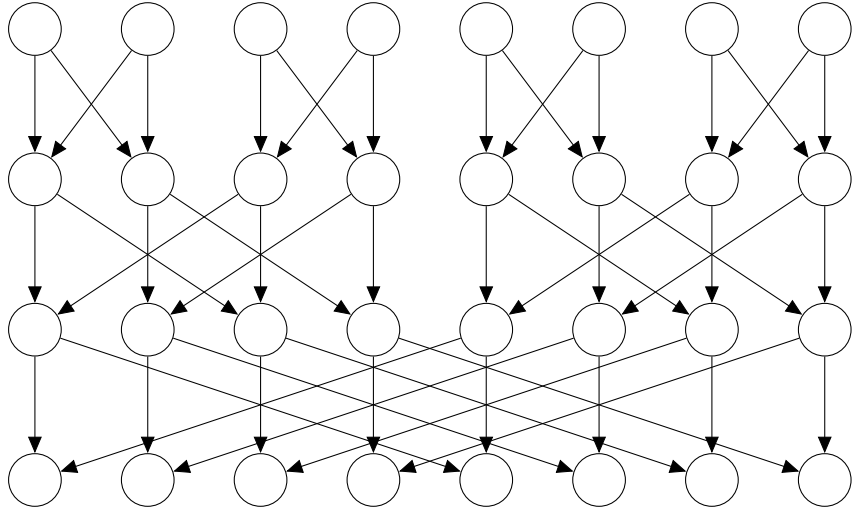

It is well known that the butterfly graph is singly-connected (see Figure \ref{fig:ExpButterflyGraph}). Actually, every source is connected to every sink via a unique path. This implies that all DFS calls on sources need time $\Theta(t)$, hence the total time is $\Theta(t^{2})$, much bigger than $m$. (but still smaller than $n^{2}=2^{2d}(d+1)^{2}$).

On Butterfly graphs our algorithm takes much more than $O(m)$ time. It remains open to find an algorithm that breaks the $O(s\cdot t)$ bound.
\section{Optimization problems related to uni-connectivity} \label{OPsRtoUniC}
In this section we study two optimization problems related to single connectivity. The first problem is defined as follows. Given a directed graph $G=(V,E)$, find an edge set $C\subseteq E$ of minimum size such that the directed graph $(V,E\setminus C)$ is singly-connected. This problem is denoted by ESC. The second problem is defined as follows. Given a directed graph $G=(V,E)$, find a minimum cardinality vertex set $F\subseteq V$ such that the directed graph $G\setminus F$ obtained from $G$ by removing all the vertices in $F$ and their incident edges is singly-connected. We denote this problem by VSC.
We show that VSC and ESC are NP-hard by reducing the vertex cover problem to each of them.

The decision version of the vertex cover problem is NP-complete \cite{K72}. It is defined as follows. Given an undirected graph $G=(V,E)$ and an integer $l$. Is there a vertex set $U\subseteq V$ with $|U|\leq l$ such that for every edge $e=(v,w) \in E$ we have $\lbrace v,w\rbrace \cap U\neq\emptyset$?

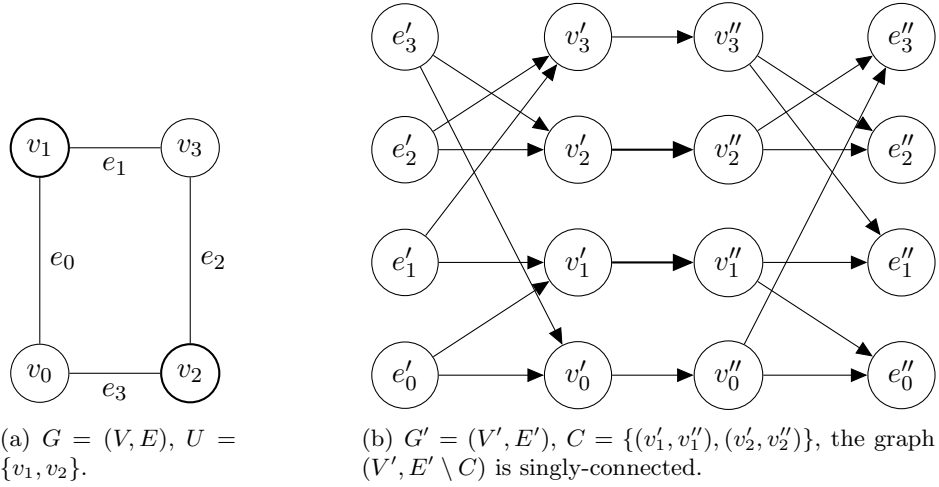
\begin{figure}[htbp]
    \centering
    \subfigure[$G=(V,E)$, $U=\lbrace v_1,v_2 \rbrace$.]{
    \begin{tikzpicture}
        \tikzstyle{every node}=[color=black,draw,circle,minimum size=0.7cm]
        \node  (v0) at (0,0) {$v_0$};
        \node [thick](v1) at (0,3) {$v_1$};
        \node [thick](v2) at (2,0) {$v_2$};
        \node (v3) at (2,3) {$v_3$};

       \begin{scope}
            \tikzstyle{every node}=[auto=right]
            \draw (v0) to node {$e_0$} (v1);
            \draw (v1) to node {$e_1$} (v3);
            \draw (v2) to node {$e_2$} (v3);
            \draw (v0) to node {$e_3$} (v2);
            
        \end{scope}
    \end{tikzpicture}} \qquad\qquad
    \subfigure[$G' =(V',E')$, $C=\lbrace (v_1',v_1'') ,(v_2',v_2'')\rbrace$, the graph $(V',E'\setminus C)$ is singly-connected.]{
    \begin{tikzpicture}
        \tikzstyle{every node}=[color=black,draw,circle,minimum size=0.7cm]
        \node (v0) at (1,0) {$v_0'$};
        \node (v1) at (1,1.5) {$v_1'$};
        \node (v2) at (1,3) {$v_2'$};
        \node (v3) at (1,4.5) {$v_3'$};
        \node (v4) at (3,0) {$v_0''$};
        \node (v5) at (3,1.5) {$v_1''$};
        \node (v6) at (3,3) {$v_2''$};
	    \node (v7) at (3,4.5) {$v_3''$};
	    \node (v8) at (-1.3,0) {$e_0'$};
        \node (v9) at (-1.3,1.5) {$e_1'$};
        \node (v10) at (-1.3,3) {$e_2'$};
        \node (v11) at (-1.3,4.5) {$e_3'$};
        \node (v13) at (5.3,0) {$e_0''$};
        \node (v14) at (5.3,1.5) {$e_1''$};
        \node (v15) at (5.3,3) {$e_2''$};
	    \node (v16) at (5.3,4.5) {$e_3''$};

       \begin{scope}
            \tikzstyle{every node}=[auto=right]
            \draw [-triangle 45](v0) to (v4);
            \draw [-triangle 45][thick](v1) to (v5);
            \draw [-triangle 45][thick](v2) to (v6);
            \draw [-triangle 45](v3) to (v7);  
            \draw [-triangle 45](v8) to (v0); 
            \draw [-triangle 45](v8) to (v1);   
            \draw [-triangle 45](v4) to (v13); 
            \draw [-triangle 45](v5) to (v13); 
            \draw [-triangle 45](v5) to (v14); 
            \draw [-triangle 45](v7) to (v14);
            \draw [-triangle 45](v9) to (v1); 
            \draw [-triangle 45](v9) to (v3);  
            \draw [-triangle 45](v10) to (v2); 
            \draw [-triangle 45](v10) to (v3); 
            \draw [-triangle 45](v6) to (v15); 
            \draw [-triangle 45](v7) to (v15); 
            \draw [-triangle 45](v11) to (v0); 
            \draw [-triangle 45](v11) to (v2); 
            \draw [-triangle 45](v4) to (v16); 
            \draw [-triangle 45](v6) to (v16); 
        \end{scope}
    \end{tikzpicture}}
   \caption{Reducing vertex cover to ESC.}
    \label{fig:figRedVCtoEUC}
\end{figure}

We define the decision version of ESC as follows: Given a directed graph $G=(V,E)$ and an integer $r$. Does there exist an edge set $C\subseteq E$ of size at most $r$ such that the directed graph $(V,E\setminus C)$ is singly-connected?
\begin{lemma} \label{def:EUCIsNPComplete}
The decision version of ESC is NP-complete.
\end{lemma}
\emph{Proof}:
It is obvious that ESC is in NP.
Let $(G=(V,E),l)$ be an instance of the vertex cover problem. We construct an instance $(G' =(V',E'),r)$ of ESC as follows. For each vertex $v\in V$, we add two new vertices $v',v''$ to $V'$ and a directed edge $(v',v'')$ to $E'$. Furthermore, for every undirected edge $e=(v,w)\in E$ we add two vertices $e',e''$ to $V'$ and four directed edges $(e',v'),(e',w'),(v'',e''),(w'',e'')$ to $E'$. An example is illustrated in Figure \ref{fig:figRedVCtoEUC}. 
Clearly, the directed graph $G'$ has $2|V|+2|E|$ vertices and $|V|+4|E|$ edges. Therefore, $G'$ can be constructed from $G$ in polynomial time. 
Now we prove that $G$ has a vertex cover of size at most $l$ if and only if $G'$ has an edge set $C\subseteq E'$ of size at most $r=l$ such that $(V',E'\setminus C)$ is singly-connected.\\
``$\Rightarrow$'': Let $U$ be a vertex cover of $G$ such that $|U|\leq l$. Let $C=\lbrace (v',v'')\mid  (v',v'') \in E' $ and $ v\in U\rbrace$. Obviously, $|C|\leq l$. For every pair of vertices $e',e'' \in V'$ which correspond to undirected edge $e=(u,w)$ in $G$, there exist two simple paths $(e',u',u'',e'')$, $(e',w',w'',e'')$ from $e'$ to $e''$ in $G'$. Since $\lbrace u,w\rbrace\cap U\neq\emptyset$, we have $\lbrace(u',u''),(w',w'') \rbrace \cap C \neq \emptyset$. Therefore, there is at most one simple path from $e'$ to $e''$ in $(V',E'\setminus C)$. Moreover, for any distinct vertices $x,y$ with $x\neq e'$ or $y\neq  e''$, there is at most one simple path from $x$ to $y$ in $G'$. Consequently, the directed graph $(V',E'\setminus C)$ is singly-connected.\\
``$\Leftarrow$'': Let $C$ be a subset of $E'$ with $|C|\leq l$ such that $(V',E' \setminus C)$ is singly-connected. Let $U=\lbrace v\mid \lbrace(e',v'), (v',v''), (v'',e'')\rbrace \cap C \neq\emptyset$ and $ v\in V \rbrace$. It is easy to see that $|U|\leq l$. Assume for a contradiction that $U$ is not a vertex cover of $G$. Then there are two vertices $u,w$ in $G$ such that $\lbrace u,w\rbrace \cap U=\emptyset$. This implies that $\lbrace(e',u'),(u',u''),(u'',e''),(e',w'),(w',w''),(w'',e'')\rbrace \cap C=\emptyset$. Thus, there are two simple paths $(e',u',u'',e'')$ and $(e',w',w'',e'')$ from $e'$ to $e''$ in the directed graph $(V',E' \setminus C)$, a contradiction. 

\begin{figure}[htbp]
    \centering
    \subfigure[$G=(V,E)$, $U=\lbrace v_1,v_2 \rbrace$ is a vertex cover of $G$.]{
    \begin{tikzpicture}
        \tikzstyle{every node}=[color=black,draw,circle,minimum size=0.7cm]
        \node  (v0) at (0,0) {$v_0$};
        \node [thick](v1) at (0,3) {$v_1$};
        \node [thick](v2) at (3.5,0) {$v_2$};
        \node (v3) at (3.5,3) {$v_3$};

       \begin{scope}
            \tikzstyle{every node}=[auto=right]
            \draw (v0) to node {$e_0$} (v1);
            \draw (v1) to node {$e_1$} (v3);
            \draw (v2) to node {$e_2$} (v3);
            \draw (v0) to node {$e_3$} (v2);
            
        \end{scope}
    \end{tikzpicture}} \qquad\qquad
    \subfigure[$G' =(V',E')$, $F=\lbrace v_1' ,v_2' \rbrace$, the graph $G'\setminus F$ is singly-connected.]{
    \begin{tikzpicture}
        \tikzstyle{every node}=[color=black,draw,circle,minimum size=0.7cm]
        \node (v0) at (1,0) {$v_0'$};
        \node [thick](v1) at (1,1.5) {$v_1'$};
        \node [thick](v2) at (1,3) {$v_2'$};
        \node (v3) at (1,4.5) {$v_3'$};
	    \node (v8) at (-1.5,0) {$e_0'$};
        \node (v9) at (-1.5,1.5) {$e_1'$};
        \node (v10) at (-1.5,3) {$e_2'$};
        \node (v11) at (-1.5,4.5) {$e_3'$};
        \node (v13) at (3.5,0) {$e_0''$};
        \node (v14) at (3.5,1.5) {$e_1''$};
        \node (v15) at (3.5,3) {$e_2''$};
	    \node (v16) at (3.5,4.5) {$e_3''$};
       \begin{scope}
            \tikzstyle{every node}=[auto=right] 
            \draw [-triangle 45](v8) to (v0); 
            \draw [-triangle 45](v8) to (v1);   
            \draw [-triangle 45](v0) to (v13); 
            \draw [-triangle 45](v1) to (v13); 
            \draw [-triangle 45](v1) to (v14); 
            \draw [-triangle 45](v3) to (v14);
            \draw [-triangle 45](v9) to (v1); 
            \draw [-triangle 45](v9) to (v3);  
            \draw [-triangle 45](v10) to (v2); 
            \draw [-triangle 45](v10) to (v3); 
            \draw [-triangle 45](v2) to (v15); 
            \draw [-triangle 45](v3) to (v15); 
            \draw [-triangle 45](v11) to (v0); 
            \draw [-triangle 45](v11) to (v2); 
            \draw [-triangle 45](v0) to (v16); 
            \draw [-triangle 45](v2) to (v16); 
        \end{scope}
    \end{tikzpicture}}
   \caption{Reducing vertex cover to VSC.}
    \label{fig:figReducingVCtoVUC}
\end{figure}
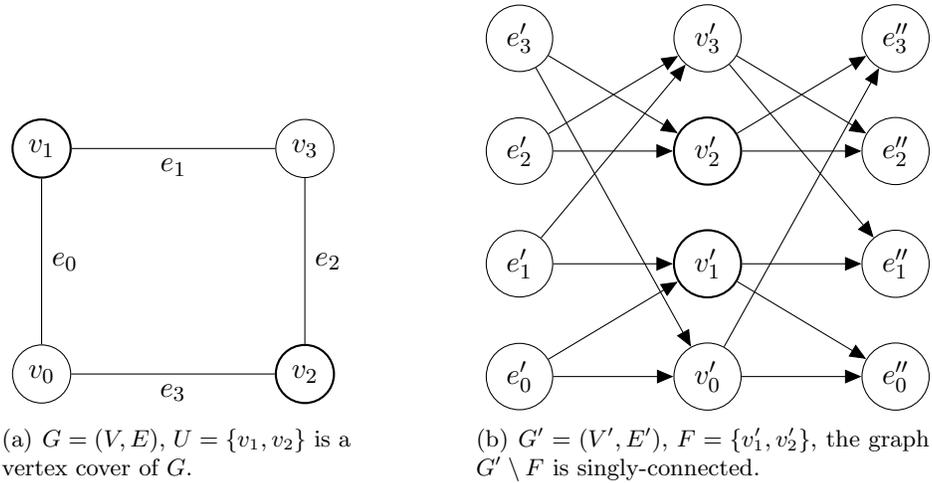

Now we define the decision version of VSC. Given a directed graph $G=(V,E)$ and an integer $r$. Is there a vertex set $F\subseteq V$ of size at most $r$ such that the directed graph $G\setminus F$ is singly-connected. We have the following.
\begin{lemma}
The decision version of VSC is NP-complete.
\end{lemma}
\emph{Proof}: The proof is similar to the proof of Lemma \ref{def:EUCIsNPComplete} (see Figure \ref{fig:figReducingVCtoVUC}).

\section{Open Problems}
 We leave as an open problem whether Karlin's solution \cite{K95}   (also see \cite{K00}) for the decision problem can be improved. 
Another open problem is whether there are approximation algorithms for the problems described in Section \ref{OPsRtoUniC}.

\end{document}